\pdfoutput=1
\documentclass[10pt,twocolumn]{article}
\usepackage[a4paper, margin=1.3cm]{geometry} 
\usepackage{indentfirst}
\usepackage[utf8]{inputenc}
\usepackage[english]{babel}
\usepackage{csquotes}
\usepackage{amsmath}
\usepackage{amsfonts}
\usepackage{amssymb}
\usepackage[per-mode=symbol]{siunitx}
\DeclareSIUnit\angstrom{\text{Å}}
\usepackage{bm}                       
\usepackage{mathpazo}                 
\usepackage{microtype}                
\usepackage{tikz}
\usepackage{graphicx}
\usepackage{float}                    
\usepackage[export]{adjustbox}        
\usepackage{flushend}                 
\usepackage[font=small,labelfont=bf]{caption}
\usepackage{subfig}
\usepackage{ifthen}                   
\usepackage{booktabs}                 
\usepackage{multirow}                 
\usepackage[symbol,hang,flushmargin]{footmisc} 
\usepackage[version=4]{mhchem}        

\usepackage{hyperref}
\hypersetup{colorlinks=true, linkcolor=blue, citecolor=blue, urlcolor=blue} 

\usepackage{scrextend}                
\changefontsizes{9pt}
\usepackage{xcolor}
\definecolor{blue1}{RGB}{ 7,  47,  95}
\definecolor{blue2}{RGB}{18,  97, 160}
\definecolor{blue3}{RGB}{56, 149, 211}
\definecolor{red}{RGB}{210, 0, 0}

\usepackage{titlesec}                 
\titlelabel{\thetitle.\ }             
\titleformat*{\section}{\color{blue1}\scshape\bfseries\centering\large}
\titleformat*{\subsection}{\color{blue2}\normalfont\itshape\large}
\titleformat*{\subsubsection}{\color{blue3}\normalfont\itshape}
\titleformat{\paragraph}[runin]{\color{blue3}\normalfont\itshape}{}{0em}{}[~--]
\titlespacing{\paragraph}{0em}{0em}{0.3em}

\usepackage[backend=bibtex,date=year,doi=false,eprint=false,giveninits=true,isbn=false,maxnames=2,maxbibnames=99,sorting=none,style=nature]{biblatex}
\AtEveryBibitem{
	\clearlist{language}
}
\newlength{\spc} 
\let\footnoteorig\footnote
\renewcommand{\footnote}[2]{
	\ifthenelse{\equal{#2}{,}\OR\equal{#2}{.}}{%
		\settowidth{\spc}{#2}
		\addtolength{\spc}{-1.8\spc}
		#2
		\hspace*{\spc}
		\footnoteorig{#1}
	}{%
		\footnoteorig{#1}%
		\ifthenelse{\NOT\equal{#2}{;}\AND\NOT\equal{#2}{:}}{\ }{}%
		#2%
	}%
} 

\renewcommand{\textcite}[3][0.15em]{\citeauthor{#2}\hspace*{-#1}\cite{#2}{#3}}
\renewcommand{\cite}[2]{
	\ifthenelse{\equal{#2}{,}\OR\equal{#2}{.}}{%
		\settowidth{\spc}{#2}
		\addtolength{\spc}{-1.8\spc}
		#2
		\hspace*{\spc}
		\supercite{#1}
	}{%
		\supercite{#1}%
		\ifthenelse{\NOT\equal{#2}{;}\AND\NOT\equal{#2}{:}\AND\NOT\equal{#2}{)}}{\ }{}%
		#2%
	}%
}

\newcommand{\snspace}[2][0.45em]{
	\hspace*{-#1}#2\hspace{0.2em}}

\begin{document}

\twocolumn[
	\begin{@twocolumnfalse}

		\begin{center}
			\textbf{\color{blue1}\large Role of minimum adhesive wear particle size in third-body layer properties}\\
			\vspace{1em}
			Son Pham-Ba\footnotemark\hspace{-0.3em},\hspace{0.1em} Jean-François Molinari\\\vspace{0.5em}
			\textit{\footnotesize Institute of Civil Engineering, Institute of Materials Science and Engineering,\\\vspace{-0.2em}
			École polytechnique fédérale de Lausanne (EPFL), CH 1015 Lausanne, Switzerland}
		\end{center}


		\begin{center}
			\parbox{14cm}{\small
				\setlength\parindent{1em}We employ a novel discrete element method (DEM) force formulation to simulate adhesive wear and assess the effects of material and loading parameters on the properties of the third-body layer (TBL) formed during sliding motion. The study emphasizes the role of a material's critical length scale $d^*$ in the rheology of the TBL. This critical length scale is already known for controlling the size of smallest wear particles. We observe the emergence of a several wear regimes involving wear particle creation and aggregation, with limited effect from $d^*$ on TBL properties. Instead, material strength and surface energy have a profound influence. This study opens up new avenues for exploration of larger systems, three-dimensional setups, and other loading conditions.

				\vspace{1em}
			}
		\end{center}

		\vspace{1em}

	\end{@twocolumnfalse}
]

\footnotetext{Corresponding author. E-mail address: \href{mailto:son.phamba@epfl.ch}{son.phamba@epfl.ch}}

\section{Introduction}

Atomic-scale studies of the phenomenon of adhesive wear\cite{aghababaeiCriticalLengthScale2016,aghababaeiDebrislevelOriginsAdhesive2017,zhaoAdhesiveWearLaw2020,aghababaeiMicromechanicsMaterialDetachment2021} suggest that the size of the smallest wear particles is of the order of a critical length scale\footnote{Also referred as the critical junction size. It has the same scaling as the fracture process zone size.} of the material
\begin{equation}
	d^* \sim \frac{\gamma E}{\sigma_\text{m}^2} \,, \label{eq:dstar_agha}
\end{equation}
where $E$ is the Young's modulus, $\gamma$ the surface energy, and $\sigma_\text{m}$ the strength of the interfaces. While the smallest wear particles contribute to the formation of a third-body layer (TBL)\cite{pham-baCreationEvolutionRoughness2021,wirthFundamentalTribochemicalStudy1994,meierhoferThirdBodyLayer2014,scholzWearGougeFormation1987,wilsonParticleSizeEnergetics2005}, there is no mention in the literature of a direct link between the material's critical length scale and the TBL characteristics (thickness, rheology, \emph{etc.}).

Experimental and numerical studies typically treat TBL and gouges macroscopically as (granular) fluids\cite{chenPowderRollingMechanism2017,pandeNumericalMethodsRock1990,cundallComputerModelSimulating1971}, whereas numerical studies at the atomic scale on comparatively very short durations clearly showcase the formation of distinct wear particles of various sizes and their aggregation\cite{aghababaeiCriticalLengthScale2016,brinkEffectWearParticles2022}. Investigating the transition between scales and behaviors is a challenging task, both for experimentalists and numericians. In experiments, the transition happens at durations that are short and challenging to observe\cite{pham-baCreationEvolutionRoughness2021}. Numerical simulations of wear, typically performed using molecular dynamics (MD), are more suited for assessing small and short events. However, the computational cost becomes too high for larger systems and longer times (\emph{e.g.} \textcite{brinkEffectWearParticles2022}). Thankfully, a new formulation of inter-particular forces was developed to tackle this particular problem using the discrete element method (DEM)\cite{pham-baAdhesiveWearCoarsegrained2022a}, while staying close to the physics achievable with MD. This model is capable of reproducing the single-asperity wear behaviors captured by MD while bringing the computational cost down by 5 orders of magnitude. This significant gain is possible thanks to a coarse-graining procedure, by modeling deformable material with discrete particles 10 times larger than atoms.\footnote{Therefore, the number of particles is divided by 1000. The time step is also allowed to be 10 times larger. Furthermore, the DEM pair force is less complex and computationally expensive than some MD potentials that mimic realistic materials.} This new method opens the door to studying TBL formation and evolution using adhesive wear physics.

While the aforementioned method was not yet used in a context involving more than the creation of a single wear particle, other numerical ways of investigating the evolution of third body were developed. For example, \textcite{zhangSignificanceThirdBody2020} and \textcite[0em]{mollonSoftDiscreteElementMethod2021} use a discrete element (DE) model with deformable grains to probe the role of discrete particle properties (\emph{e.g.} stiffness and friction coefficient) on the overall physical behavior of the sliding interface. Mollon's work\cite{mollonSolidFlowRegimes2019} showcases the emergence of several types of microstructures, linked to different flow regimes. As it is commonly accepted with DEM simulations, the parameters given to the particles are not directly linked to material or loading parameters. The choice of the size of the particles is also not necessarily physically motivated.

The coarse-grained DE model of \textcite[0em]{pham-baAdhesiveWearCoarsegrained2022a} allows performing similar simulations while directly linking the physics of the system to material properties. Systems studied using this model are shown to produce wear particles having a size independent of the size of the discrete elements (in a given range) and controlled instead by material properties and geometrical parameters. This feature removes the concern about the impact of the size of the DE (which has not always physical meaning) on the behavior of the whole system.

In this exploratory study, we perform numerical simulations of adhesive wear under sliding motion using the coarse-grained DE method\cite{pham-baAdhesiveWearCoarsegrained2022a}. Several material and loading parameters are evaluated, and the effects on the properties of the formed third-body layer are assessed, putting an emphasis on the role of the material's critical length scale $d^*$ on the rheology of the TBL.

\section{Method}

The chosen numerical method allows us to simulate a dynamical tribological system where detachment and reattachment of matter can occur, using spherical DE with cohesive forces. The inter-particular force parameters can be chosen to match several material parameters, notably the Young's modulus, the strength, and the surface energy.

\begin{figure}
	\centering
	\includegraphics{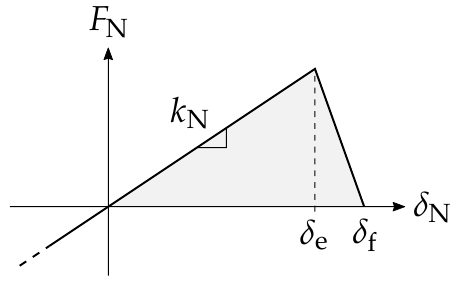}
	\caption[Normal force between two particles as a function of inter-particular distance $\delta_\text{N}$]{Normal force between two particles as a function of inter-particular distance $\delta_\text{N}$. There is interpenetration when $\delta_\text{N} < 0$. The force has a cohesive part when $\delta_\text{N} > 0$.}
	\label{fig:dem_forces}
\end{figure}

The force $\bm{F}$ acting between a given pair of particles acts along the line going through the center of the particles (the normal direction, relative to particles' surface). It is the sum of a stiffness or cohesive component $F_\text{N}$ and a velocity damping force:
\begin{equation}
	\bm{F} = -(F_\text{N} + c_\text{N}v_\text{N}) \bm{n}_\text{N} \,,
\end{equation}
where $\bm{n}_\text{N}$ is the unit vector pointing in the normal direction, $v_\text{N}$ is the corresponding relative velocity, and $c_\text{N}$ is a damping factor. The formulation used here is simplified compared to the full formulation established by \textcite[0em]{pham-baAdhesiveWearCoarsegrained2022a}. The tangential forces between the particles are not modeled, reducing the number of model parameters, the trade-off being that the target Poisson's ratio of the modeled material is fixed at $\nu = 0.25$. Macroscopic apparent shear forces still exist since the normal inter-particular forces are sufficiently long-ranged. The expression of the normal force is
\begin{equation}\label{eq:FN}
	F_\text{N} = \begin{cases}
		k_\text{N} \delta_\text{N} & \text{if } \delta_\text{N} \leqslant \delta_\text{e} \,, \\
		\displaystyle -\frac{k_\text{N} \delta_\text{e}}{\delta_\text{f} - \delta_\text{e}} (\delta_\text{N} - \delta_\text{f}) & \text{if } \delta_\text{e} < \delta_\text{N} \leqslant \delta_\text{f} \,, \\
		0 & \text{otherwise,}
	\end{cases}
\end{equation}
where $\delta_\text{N}$ is the distance between the particles' surfaces, $\delta_\text{e}$ is the maximum elastic distance, $\delta_\text{f}$ is the fracture distance, and $k_\text{N}$ is the Hookean stiffness. The profile of the force is shown in Figure~\ref{fig:dem_forces}.

The stiffness, distance, and damping parameters are linked to the desired material properties through the following equations:
\begin{align}
	&k_\text{N} = \frac{A_\text{eff}E}{r_i + r_j} \,, \label{eq:kN} \\
	&\delta_\text{e} = \frac{(r_i + r_j) \sigma_\text{m}}{E} \,, \\
	&\delta_\text{f} = \displaystyle \frac{4\gamma}{\sigma_\text{m}} \label{eq:df} \,, \\
	&c_\text{N} = \frac{2(1 - \eta)}{\pi} \sqrt{k_\text{N}m_\text{eff}} \,, \label{eq:cN} \\
	&r_\text{eff} = \min(r_i, r_j) \,, \\
	&A_\text{eff} = 2\sqrt{2} \, r_\text{eff}^2 \,, \\
	&m_\text{eff} = \frac{m_i m_j}{m_i + m_j} \,,
\end{align}
where $r_i$ and $r_j$ are the radii of the interacting particles, $m_i$ and $m_j$ their respective masses, $E$ is the target Young's modulus, $\sigma_\text{m}$ the target strength, $\gamma$ the target surface energy, and $\eta$ the target restitution coefficient. These equations are valid as long as the radii of the particles satisfy
\begin{equation}
	r_i + r_j \leqslant d_\text{m} \,, \label{eq:r_limit}
\end{equation}
where
\begin{equation}
	d_\text{m} = 4 \frac{\gamma E}{\sigma_\text{m}^2} \label{eq:dmax}
\end{equation}
can be seen as a maximum discrete particle diameter (specific to this formulation but still closely related to the material's critical length scale $d^*$ \eqref{eq:dstar_agha}). If \eqref{eq:r_limit} is not satisfied, the actual values of $\sigma_\text{m}$ and $\gamma$ (of the DEM system) may deviate from their imposed targets.

The Young's modulus $E$, the density $\rho$, and the size of the system $L$ are all arbitrarily set to $1$ (a coherent system of units can also be chosen arbitrarily). Their actual value is not important, as all the other dimensional quantities can be expressed relative to those three. The restitution coefficient between particles (dimensionless) is set to $\eta = 0.95$.\footnote{While this value may seem high for DEM practitioners, it actually makes sense to have a low amount of damping at this near atomic scale ($d_0$ around $10$ times larger than atoms that would make a material with such properties).} Thus, the only remaining free parameters needed to define the material are the strength $\sigma_\text{m}$ and the surface energy $\gamma$.

Since our goal is to relate the material parameters to the properties of the TBL, such as its thickness, it is sensible to combine the material parameters into a quantity that has units of length. We can use the critical length of the material \eqref{eq:dstar_agha}, which was found numerically\footnote{A system comprised of two surfaces linked by a square junction of size $d$ is sheared at a constant rate, without normal load. The critical size $d^*$ is identified such that junctions with a size $d < d^*$ flow plastically, and those with $d \geqslant d^*$ detach into a wear particle.} in our situation to be roughly
\begin{equation}
	d^* = 32 \frac{\gamma E}{\sigma_\text{m}^2} \,, \label{eq:dstar}
\end{equation}
following the scaling of \eqref{eq:dstar_agha}, with the interface strength taken as the material bulk tensile strength $\sigma_\text{m}$. The last parameter of interest is the normal pressure $p_\text{N}$ applied on the system. Our simulations differ from those in \textcite[0em]{mollonSolidFlowRegimes2019}, where the gap between the surfaces is kept constant instead of imposing a normal load.

\begin{figure}
	\centering
	\includegraphics{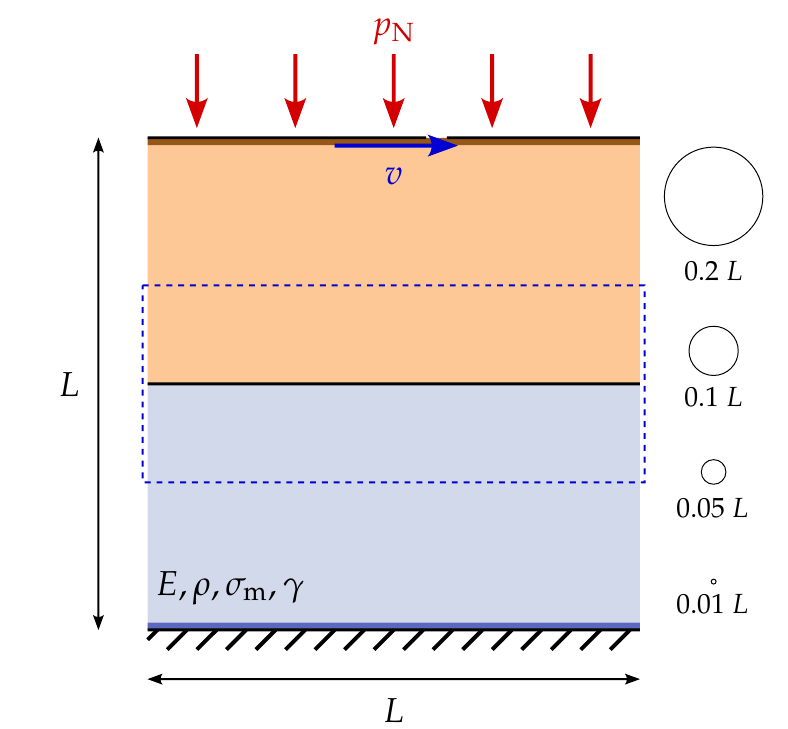}
	\caption[Schematic of the simulated DEM systems]{Schematic of the simulated DEM systems. On the right part of the figure, some relevant sizes are represented. The whole system is filled with particles whose mean diameter is $d_0 = 0.01\ L$. The thickness of the system (quasi two-dimensional) is $B = 3\ d_0$. The material properties are chosen such that the resulting $d^*$ is equal to either $0.05\ L$, $0.1\ L$, or $0.2\ L$, which are represented on the right. The thin darker areas at the top and bottom of the system are modeled as rigid bodies. The particles making them cannot move relative to each other. The bottom rigid body is fixed, while the top one receive a constant compressive normal pressure $p_\text{N}$ and is dragged horizontally at a constant velocity $v$. At the initial state of the simulations, the top and bottom part of the system are disjoined, \emph{i.e.} there is no particle that is part of both the orange body and the blue body. The area enclosed by the dashed line is the region represented in the subsequent figures.}
	\label{fig:system}
\end{figure}

All studied systems fit in a quasi-2D square of side-length $L$ (see Figure~\ref{fig:system}). The thickness $B$ of the system is adjusted to the average size $d_0 = 0.01 \, L$ of the particles to be $B = 3 \, d_0$. Using such a system instead of a fully two-dimensional one allows us to use spherical particles with sizes distributed around a mean value to create a disordered system without weak planes, while ensuring a tight packing of the particles. The sizes of the particles follow a log-normal distribution\footnote{In this case, with the truncation, the distribution is similar to a Gaussian distribution.} with a mode (most frequent value) of $d_0$ and a standard deviation of $0.1 \, d_0$. The distribution is truncated between $0.75 \, d_0$ and $1.25 \, d_0$. It is shown in \textcite[0em]{pham-baAdhesiveWearCoarsegrained2022a} that $d_0$ and the whole distribution can be varied without significantly changing the properties of the resulting medium or its adhesive wear behavior, as long as $d_0$ is smaller than $d^*$ (the minimum wear particle size for the target material). An amorphous arrangement of particles is generated by first inserting the particles at random positions into the system, then performing a critically damped dynamical simulation, which makes the particles rearrange into a configuration that minimizes overlaps. The boundaries of the system are allowed to move such that the average internal stresses drop to $0$. More details regarding the exact parameters of the relaxation procedure are given by \textcite[0em]{pham-baAdhesiveWearCoarsegrained2022a}.

The simulated systems consist of two halves filling the space of size $L \times L \times B$ and having a non-matching interface (in the sense of the discrete particles' arrangement). The bottom half is connected to a fixed rigid body (made of non-moving particles) of height $1.5 \, d_0$, and the top half is dragged by a rigid body of the same size at a fixed velocity of $0.1 \sqrt{E / \rho}$ over a total distance of $50 \, L$. The other directions have periodic boundary conditions. A normal pressure $p_\text{N}$ is applied on the top rigid body.

The normal pressure is varied between $0.01 \, E$, $0.02 \, E$, and $0.04 \, E$. For the other two parameters, one could boldly expect the thickness of the TBL to be only dependent on the value of $d^*$ (linked to the minimum size of wear particles), but not on the individual values of $\gamma$ and $\sigma_\text{m}$. So, at first, to assess the direct influence of $d^*$\snspace, the strength is set to $\sigma_\text{m} = 0.1 \, E$ and the surface energy $\gamma$ is adjusted using \eqref{eq:dstar} to reach $d^*$ values of $0.05 \, L$, $0.1 \, L$, and $0.2 \, L$. Then, in a second stage, to assess the individual effects of $\sigma_\text{m}$ and $\gamma$, the critical size is fixed to $d^* = 0.1 \, L$ and the strength is varied between $0.1 \, E$, $0.14 \, E$, and $0.2 \, E$, while the surface energy $\gamma$ is still adjusted (increased when $\sigma_\text{m}$ increases) to match $d^*$\snspace. In summary, for each normal pressure $p_\text{N}$, the tested sets of $\sigma_\text{m}$ and $\gamma$ are shown in Table~\ref{tab:param_sets}. To get a sense of how material parameters affect the interaction between particles, we can also compute the maximum interaction distance $\delta_\text{f}$ given by \eqref{eq:df}, which is related to ductility.

\begin{table}[H]
	\centering
	\caption{Sets of tested parameters (apart from $p_\text{N}$) and particles' maximum interaction distance}
	\label{tab:param_sets}
	\small
\begin{tabular}{ccc|c}
	\toprule
	$d^*$ & $\sigma_\text{m}$ & $\gamma$ & $\delta_\text{f}$ \\
	\midrule
	$0.05 \, L$ & $0.1  \, E$ & $1.56 \cdot 10^{-5} \, EL$ & $0.06 \, d_0$ \\
	$0.1  \, L$ & $0.1  \, E$ & $3.12 \cdot 10^{-5} \, EL$ & $0.12 \, d_0$ \\
	$0.2  \, L$ & $0.1  \, E$ & $6.25 \cdot 10^{-5} \, EL$ & $0.25 \, d_0$ \\
	\midrule
	$0.1  \, L$ & $0.1  \, E$ & $3.12 \cdot 10^{-5} \, EL$ & $0.12 \, d_0$ \\
	$0.1  \, L$ & $0.14 \, E$ & $6.12 \cdot 10^{-5} \, EL$ & $0.17 \, d_0$ \\
	$0.1  \, L$ & $0.2  \, E$ & $12.5 \cdot 10^{-5} \, EL$ & $0.25 \, d_0$ \\
	\bottomrule
\end{tabular}

\end{table}

\section{Results and discussion}

All the performed simulations can be qualitatively categorized into one of the three aspects shown in Figure~\ref{fig:qual} and in Supplemental videos. In the first case \subref{subfig:qual_mixed}, wear particles of various sizes are formed, continuously agglomerated together, and reattached to the sliding bodies. The size of the gap remains visually stationary. In the second aspect \subref{subfig:qual_band}, comparable to a fluid flow, the two bodies are fully connected and some mixing (vertical diffusion) occurs over time. In some instances, the simulations transition into another state \subref{subfig:qual_particle}, featuring large wear particles, growing in size, and reaching up to the full size of the system. In any case, there is always a first stationary phase, \subref{subfig:qual_mixed} or \subref{subfig:qual_band}, lasting for an arbitrary amount of time. In the quantitative analysis of the simulations, reported thereafter, we dropped the ending parts involving large growing wear particles \subref{subfig:qual_particle}. It is unknown whether all systems would eventually reach this third regime, as our simulations are currently relatively limited in time (sliding distance of $50 \, L$). Also, the limited size of the system hinders the analysis, as the TBL can travel vertically and reach the system's boundary, triggering a transition toward an unphysical regime, not represented here. Therefore, there is still room for investigation about the mechanisms triggering the third regime and for inspecting the structure of the resulting TBL. We suspect that there might be specific geometric and loading requirements for the transition to happen, that are randomly met in some simulations. For further studies, one way of simulating taller systems over longer periods of time without absurd computational costs could be to use coupling methods between finite elements and DEM\cite{voisinFiniteElementMethod2022}.

\begin{figure*}
	\centering
	\subfloat[Mixed regime\label{subfig:qual_mixed}]{
		\includegraphics{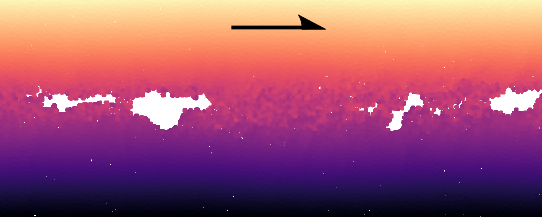}
	}
	\subfloat[Shear band\label{subfig:qual_band}]{
		\includegraphics{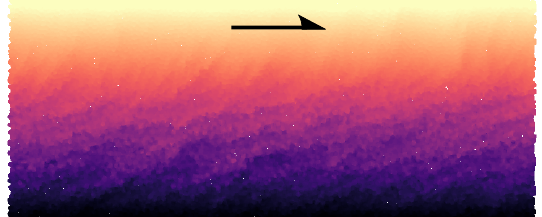}
	}
	\subfloat[Large wear particles\label{subfig:qual_particle}]{
		\includegraphics{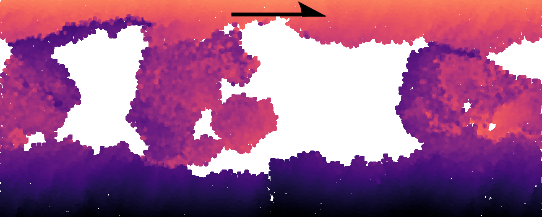}
	}
	\raisebox{-2mm}{\hspace{1mm}\includegraphics{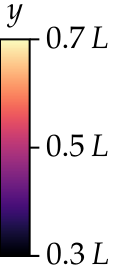}}
	\caption[Examples of qualitative regimes]{Examples of qualitative regimes. The images are cropped according to the region shown in Figure~\ref{fig:system}. Videos for each case are available as Supplementary material. \textbf{\subref{subfig:qual_mixed}} Example of mixed regime, featuring the formation of wear particles, their aggregation, and reattachment to the sliding bodies. The gap remains constant in size. The colors represent the initial vertical position of the particles. The parameters of the simulation are $p_\text{N} = 0.01 \, E$, $d^* = 0.1 \, L$, and $\sigma_\text{m} = 0.1 \, E$. \textbf{\subref{subfig:qual_band}} Example of shear band. There is no visible gap between the two bodies, and the particles are migrating vertically. The parameters of the simulation are $p_\text{N} = 0.01 \, E$ and $d^* = 0.1 \, L$, as in \subref{subfig:qual_mixed}, and $\sigma_\text{m} = 0.2 \, E$. \textbf{\subref{subfig:qual_particle}} Example of formation of large wear particles, that grow in size until reaching the size of the system. The parameters of the simulation are $p_\text{N} = 0.01 \, E$ and $d^* = 0.1 \, L$, as in \subref{subfig:qual_mixed} and \subref{subfig:qual_band}, and $\sigma_\text{m} = 0.14 \, E$.}
	\label{fig:qual}
\end{figure*}

The emergence of a stationary state featuring wear particles and reattachment (Figure~\ref{fig:qual}\subref{subfig:qual_mixed}) is worth noting, as it is not something that could be observed using MD simulations limited to a smaller scale, or larger scale DEM simulations missing the details of wear particle formation.

\begin{figure*}
	\centering
	\subfloat[Averaged horizontal velocity profile\label{subfig:vel}]{
		\includegraphics{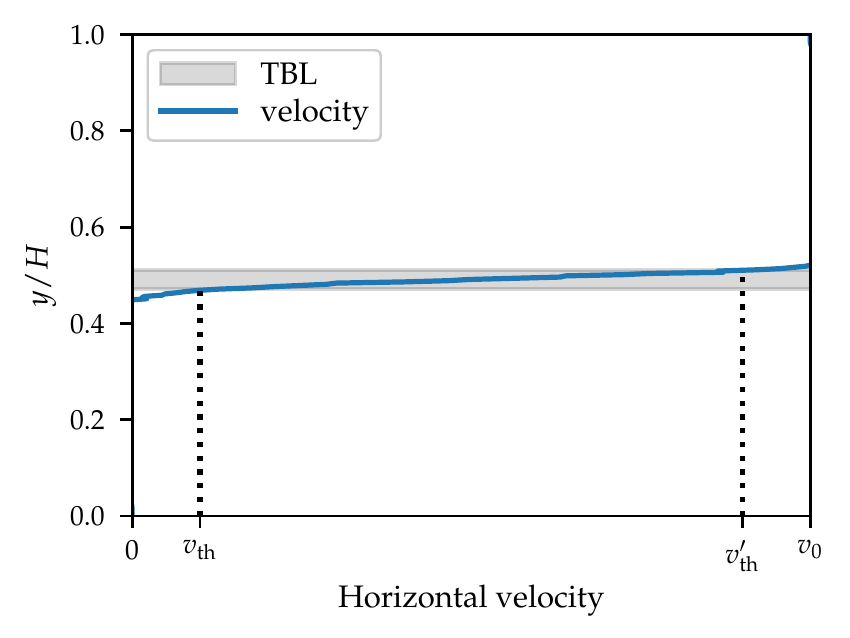}
	}
	\subfloat[Estimated density profile\label{subfig:dens}]{
		\includegraphics{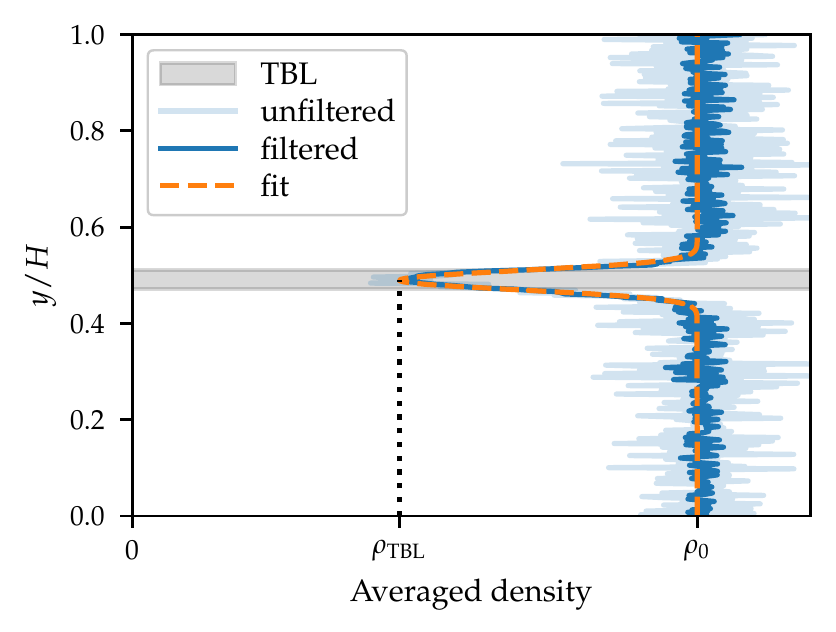}
	}
	\caption[Examples of velocity and density profiles]{Examples of velocity and density profiles. The vertical coordinate is normalized by the current height $H$ of the system (initially equal to $L$). \textbf{\subref{subfig:vel}} Example of averaged horizontal velocity profile as a function of the vertical position. The vertical axis is cut into $400$ bins and the average velocity of each section is computed. The particles have an average diameter of $d_0 = 0.01 \, L$. The boundaries of the third body layer (TBL, shaded area) are determined by the velocity profile and velocity thresholds. The bottom boundary of the system is fixed, and the top has an imposed velocity of $v_0$. Here, the thresholds are $v_\text{th} = 0.1 \, v_0$ and $v'_\text{th} = 0.9 \, v_0$. The TBL thickness $h_\text{TBL}$ can be deduced. \textbf{\subref{subfig:dens}} Example of estimated density profile. As for the velocity profile, the density is computed by binning on $400$ sectors (`filtered' curve). The density profile is filtered using a Savitzky–Golay filter of order $3$ on $7$ samples. A Gaussian curve (chosen arbitrarily) is also least-square fitted to the unfiltered data, leading to the minimum density inside the TBL $\rho_\text{TBL}$. TBL boundaries estimated from the velocity profile are shown for comparison (gray region). It is possible to have a homogeneous density with $\rho_\text{TBL} = \rho_0$, while still having well-defined TBL boundaries (given by the velocity profile).}
	\label{fig:profiles}
\end{figure*}

\begin{figure*}
    \centering
	\includegraphics{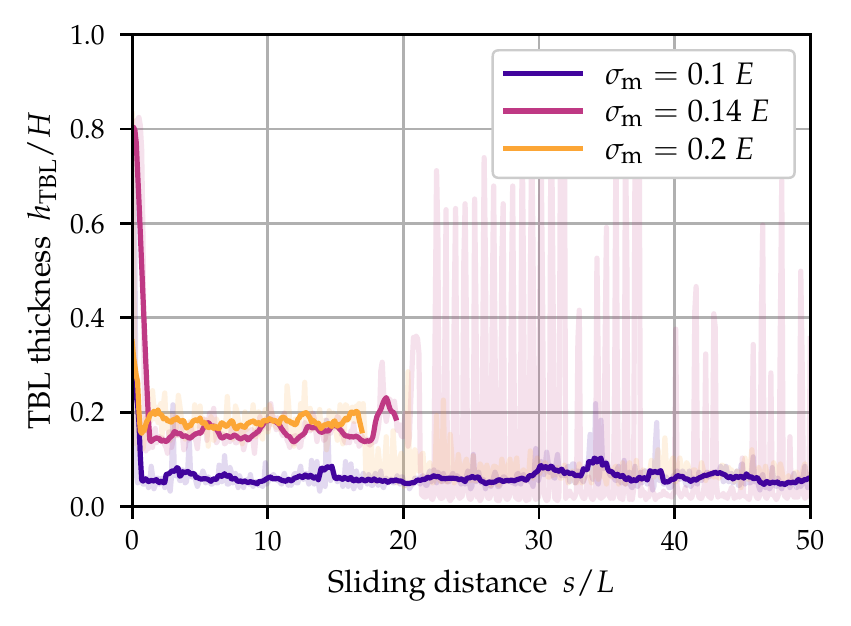}
	\caption[Examples of TBL thickness evolution]{Examples of TBL thickness evolution with $p_\text{N} = 0.04\ E$ and $d^* = 0.1\ L$. Raw measurements are shown as transparent lines, whereas the solid lines are moving averages. At the very beginning of sliding, before any amount of damage, the shear deformation is uniform, which is seen here as a peak starting from almost $h_\text{TBL} / H = 1$, $H$ being the current height of the system. A steady regime is quickly reached, being either a `mixed´ regime or a shear band. After some time, some simulations transition into the regime of large wear particles formation, identifiable here by erratic changes of TBL thickness. In other cases, the TBL reaches a boundary of the system (which is not physical), and the TBL thickness drops.}
    \label{fig:height_evo}
\end{figure*}

The geometric properties of the TBL can also be measured quantitatively. The boundaries of the TBL (and thus its thickness) are computed by looking at the horizontal velocity profile along the vertical axis (see Figure~\ref{fig:profiles}\subref{subfig:vel}). This measurement is not very accurate because the velocity profile is oscillating, meaning that arbitrary thresholds must be used to determine the boundaries of the TBL. Still, we can confirm the stationary nature of the main simulation phases by using the evolution of the TBL thickness over time (see Figure~\ref{fig:height_evo}). We notice that in the case of the shear banding (Figure~\ref{fig:qual}\subref{subfig:qual_band}), even if vertical mixing of the particles is observed over the whole height of the body, the TBL thickness (related to the vertically confined shear strain) keeps a finite size. A similar mixing behavior was observed in MD simulations\cite{kimMDSimulationsMicrostructure2007}.

The average density of particles inside the TBL can also be measured (see Figure~\ref{fig:profiles}\subref{subfig:dens}). This measurement is more reliable and accurate than the thickness measurement. When there is a visible gap between the sliding surfaces (populated by wear particles, see Figure~\ref{fig:qual}\subref{subfig:qual_mixed}), the TBL density is lower than the density of the bodies.

\begin{figure*}
	\centering
	\subfloat[Dependence on $p_\text{N}$ and $d^*$ at constant $\sigma_\text{m}$\label{subfig:dens_f_dstar}]{
		\includegraphics{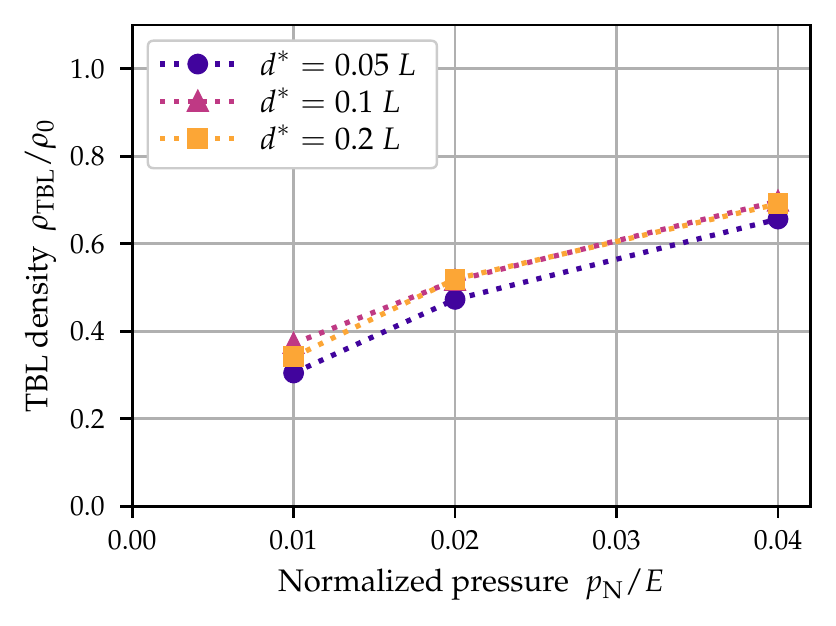}
	}
	\subfloat[Dependence on $p_\text{N}$ and $\sigma_\text{m}$ at constant $d^*$\label{subfig:dens_f_sm}]{
		\includegraphics{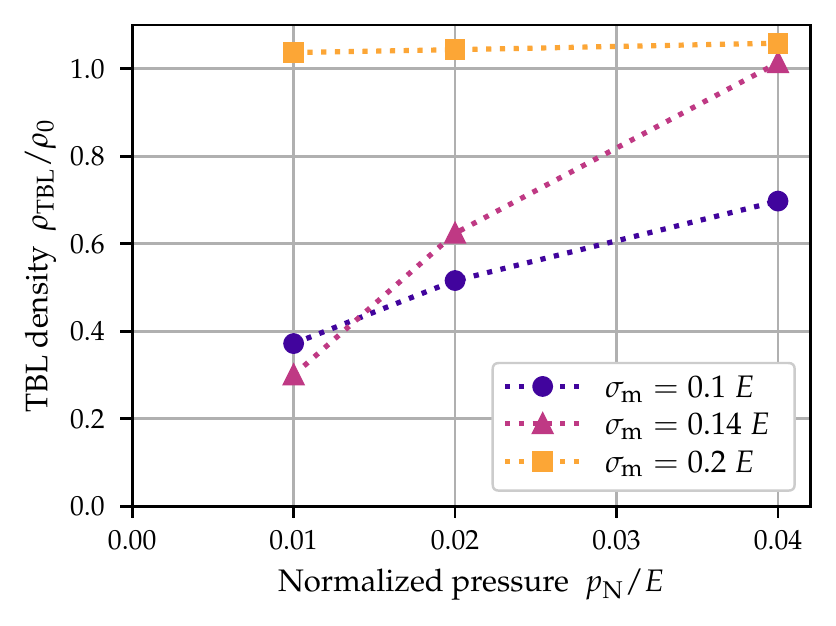}
	}
	\caption[Dependence of TBL minimum density on material properties]{Dependence of TBL minimum density on material properties. \textbf{\subref{subfig:vel}} Dependence on pressure and material's characteristic size (critical length scale). The material strength is kept constant at $\sigma_\text{m} = 0.1 \, E$ and the surface energy $\gamma$ is accommodated to obtain different values of $d^*$. The density depends mostly on the normal pressure $p_\text{N}$. \textbf{\subref{subfig:dens}} Dependence on pressure and material strength at a constant characteristic size. The surface energy $\gamma$ is increased conjointly with the strength $\sigma_\text{m}$ in order to keep the characteristic size constant at $d^* = 0.1 \, L$. Even if $d^*$ is constant, the TBL density is strongly dependent on the other parameters.}
	\label{fig:dens}
\end{figure*}

The dependence of the TBL density $\rho_\text{TBL}$ on $d^*$ and the normal pressure is shown in Figure~\ref{fig:dens}\subref{subfig:dens_f_dstar} (the strength is kept constant at $\sigma_\text{m} = 0.1 \, E$). Surprisingly, $d^*$ has no significant influence on $\rho_\text{TBL}$, compared to the effect of $p_\text{N}$. Imposing a larger normal load on the bodies has the expected effect of increasing the TBL density, which can be interpreted as increased wear volume production. The positive correlation between load and wear rate is compatible with the experimentally supported Archard's wear model\cite{archardContactRubbingFlat1953}. With these parameters (strength fixed), neither $d^*$ nor $p_\text{N}$ has a significant effect on the TBL thickness (not shown here).

Figure~\ref{fig:dens}\subref{subfig:dens_f_sm} shows the individual effects of the material parameters, where $d^*$ is kept constant. There is a clear effect of increasing the strength on the TBL density. A larger strength (and surface energy) keeps the TBL density high, meaning that the gap between the bodies remains more easily closed. The maximum density is a bit over the initial density of the material due to the normal pressure applied on the system. Increasing the strength and the surface energy (at constant $d^*$) also has the overall effect of increasing the TBL thickness, intuitively comparable to a more viscous flow.

Regarding friction, at the scale we are working, tangential forces are dominated by adhesive forces. Therefore, rather than looking at the friction coefficient $p_\text{T} / p_\text{N}$, we look directly at the average value of $p_\text{T}$ in the different steady-state regimes. $p_\text{T}$ is computed from the total horizontal force acting on the top or bottom boundary. Being mainly controlled by adhesive forces, $p_\text{T}$ is mostly dependent on the material's (tensile) strength $\sigma_\text{m}$ and on the contact area between the top and bottom bodies. When there is full contact (shear band regime shown in Figure~\ref{fig:qual}\subref{subfig:qual_band}, maximum TBL density), $p_\text{T}$ is equal to the shear strength of the material, which is measured to be roughly equal to $\tau_\text{m} = 0.22 \, \sigma_\text{m}$. In the mixed regime (Figure~\ref{fig:qual}\subref{subfig:qual_mixed}), the contact area is lower, and it depends on the normal pressure $p_\text{N}$. In fact, the contact area can be estimated to be proportional to the TBL density, and the measured $p_\text{T}$ is indeed near $\tau_\text{m} \rho_\text{TBL} / \rho_{0}$. As such, $p_\text{T}$ is not directly dependent on $d^*$ (like $\rho_\text{TBL}$). Finally, in the regime with large wear particles (Figure~\ref{fig:qual}\subref{subfig:qual_particle}), the contact area between the sliding bodies and the rolling particles is very small (not correlated with the TBL density anymore). This leads to a lubricated behavior, with tangential loads as low as $0.001 \, E$. This lubricative wear behavior is also observed experimentally\cite{zanoriaFormationCylindricalSlidingWear1995}. Understanding how it arises and how it is controlled could be very beneficial for the tribology community.

\section{Conclusion}

In this short exploratory numerical study, we demonstrated the capabilities of the coarse-grained discrete element model of \textcite[0em]{pham-baAdhesiveWearCoarsegrained2022a} in representing adhesive wear situations. We showcased the emergence of a mixed regime involving wear particle creation and their aggregation into a third-body layer. The material's critical length scale, known to play a major role in the sizing of the first formed wear particles, was shown to have a limited effect on the properties of the third-body layer. Instead, the strength and the surface energy of the material must be considered directly when treating third-body macroscopically. This works opens the door to the exploration of more ambitious setups involving adhesive wear and the formation and evolution of a third-body layer. The next steps could involve larger systems, fully three-dimensional systems, and other loading conditions such as gap-controlled sliding.

\section*{Supplementary material}

Videos for the simulations shown in Figure~\ref{fig:qual} are available along the online version of this article.

\flushcolsend 
\printbibliography
\flushcolsend

\end{document}